\documentclass[sigplan,screen,10pt]{acmart}
\copyrightyear{2023}
\acmYear{2023}
\setcopyright{licensedusgovmixed}\acmConference[PPOPP '23]{Proceedings of the 2023 Principles and Practice of Parallel Programming}{February 25-- March 01, 2023}{Montreal, Canada}
\acmBooktitle{Proceedings of the 2023 Principles and Practice of Parallel Programming, February 25-- March 01, 2023, Montreal, Canada}
\acmPrice{15.00}

\usepackage[subtle]{savetrees}

\usepackage[textsize=tiny]{todonotes} 
\usepackage[normalem]{ulem} 

\usepackage{amsmath}
\usepackage{amsthm}
\usepackage{amstext}
\usepackage{amsfonts}
\usepackage{mathrsfs}
\usepackage{xcolor}
\usepackage{tikz}
\usepackage{cleveref}
\usepackage{subcaption}
\usepackage{pgfplots}
\pgfplotsset{compat=1.3}
\usepackage{url}
\usepackage{xspace}
\usepackage{algorithm}
\usepackage[noend]{algpseudocode}
\usepackage{siunitx}
\usepackage{booktabs}
\usepackage{multirow}
\usepackage{comment}
\usepackage[title]{appendix}
\usetikzlibrary{external}
\usepackage{enumitem}
\usepackage{balance}

\usepackage{graphicx}

\usepackage{svg}

\newcommand{\errbits}{\log (1/\epsilon)}
\newcommand{\opt}{n\errbits}
\renewcommand{\epsilon}{\varepsilon}

\newcommand{\defn}[1]{{\textit{\textbf{\boldmath #1}}}\xspace}

\makeatletter
\algrenewcommand\ALG@beginalgorithmic{\footnotesize}
\makeatother

\definecolor{magenta4}{rgb}{0.5625,0,0.5625}
\definecolor{green4}{rgb}{0,0.5625,0}
\definecolor{orange4}{rgb}{0.98,0.31,0.09}
\definecolor{powderblue}{rgb}{0.69,0.88,0.9}

\newenvironment{revaddition}[1]{}

\newcommand{\revpointer}[1]{}
\newenvironment{scaddition}{}{}

\newenvironment{addition}{}{}

\newenvironment{ppoppaddition}[1]{{#1}}

\newcommand{\kathy}[1]{\todo[size=\tiny,color=powderblue]{Kathy: #1}}
\newcommand{\steve}[1]{\todo[size=\tiny,color=orange]{Steve: #1}}
\newcommand{\prashant}[1]{\todo[size=\tiny,color=cyan]{Prashant: #1}}
\newcommand{\hunter}[1]{\todo[size=\tiny,color=pink]{Hunter: #1}}

\newcommand{\kathyinline}[1]{\todo[inline,size=\tiny,color=powderblue]{Kathy: #1}}
\newcommand{\steveinline}[1]{\todo[inline,size=\tiny,color=orange]{Steve: #1}}
\newcommand{\prashantinline}[1]{\todo[inline,size=\tiny,color=cyan]{Prashant: #1}}
\newcommand{\hunterinline}[1]{\todo[inline, size=\tiny,color=pink]{Hunter: #1}}

\newcommand{\sysname}{GPU-based quotient filter\xspace}

\newcommand{\qf}{quotient filter\xspace}
\newcommand{\cqf}{counting quotient filter\xspace}
\newcommand{\bloom}{Bloom filter\xspace}

\newcommand{\para}[1]{\smallskip\noindent\textbf{#1.}}

\begin{document}

%


\newcommand{\figwidth}{0.49\textwidth}

\pgfplotsset{
  InstanPlot/.style={
    small,
    width = \linewidth + 15pt,
    height=0.7\linewidth,
    xlabel={Load factor},
    ylabel={Throughput (B/s)},
    xlabel near ticks,
    x label style={at={(0.5,-0.12)},font=\small},
    scaled y ticks = base 10:-9,
    ylabel near ticks,
    y label style={at={(-0.1,0.5)},font=\small},
    ytick style={draw=none},
    ytick scale label code/.code = {},
    yticklabel style={ /pgf/number format/fixed },
    ymajorgrids,
    yminorgrids,
    ymin=0,
    minor tick num=1,
    minor grid style={draw=gray!25},
  },
}

\pgfplotsset{
  CGPlot/.style={
    small,
    width = \linewidth + 10pt,
    height=0.6\linewidth,
    xlabel={Cooperative Group Size},
    ylabel={Throughput (B/s)},
    xlabel near ticks,
    x label style={at={(0.5,-0.12)},font=\small},
    scaled y ticks = base 10:-3,
    ylabel near ticks,
    y label style={at={(-0.1,0.5)},font=\small},
    ytick style={draw=none},
    ytick scale label code/.code = {},
    yticklabel style={ /pgf/number format/fixed },
    ymajorgrids,
    yminorgrids,
    ymin=0,
    minor tick num=1,
    minor grid style={draw=gray!25},
  },
}

\pgfplotsset{
  AggregatePlot/.style={
    small,
    width = \linewidth + 15pt,
    height=0.6\linewidth,
    xlabel={Filter Size},
    ylabel={Throughput (B/s)},
    xlabel near ticks,
    x label style={at={(0.5,-0.12)},font=\small},
    scaled y ticks = base 10:-3,
    ylabel near ticks,
    y label style={at={(-0.1,0.5)},font=\small},
    ytick style={draw=none},
    ytick scale label code/.code = {},
    yticklabel style={ /pgf/number format/fixed },
    ymajorgrids,
    yminorgrids,
    ymin=0,
    minor tick num=1,
    minor grid style={draw=gray!25},
  },
}

\pgfplotsset{
  AggregatePlotClipped/.style={
    small,
    width = \linewidth + 15pt,
    height=0.6\linewidth,
    xlabel={Filter Size},
    ylabel={Throughput (B/s)},
    xlabel near ticks,
    x label style={at={(0.5,-0.12)},font=\small},
    scaled y ticks = base 10:-3,
    ylabel near ticks,
    y label style={at={(-0.1,0.5)},font=\small},
    ytick style={draw=none},
    ytick scale label code/.code = {},
    yticklabel style={ /pgf/number format/fixed },
    ymajorgrids,
    yminorgrids,
    ymin=0,
    minor tick num=1,
    minor grid style={draw=gray!25},
  },
}

\pgfplotsset{
  AggregatePlotClippedPerl/.style={
    small,
    width = \linewidth + 15pt,
    height=0.6\linewidth,
    xlabel={Filter Size},
    ylabel={Throughput (B/s)},
    xlabel near ticks,
    x label style={at={(0.5,-0.12)},font=\small},
    scaled y ticks = base 10:-3,
    ylabel near ticks,
    y label style={at={(-0.1,0.5)},font=\small},
    ytick style={draw=none},
    ytick scale label code/.code = {},
    yticklabel style={ /pgf/number format/fixed },
    ymajorgrids,
    yminorgrids,
    ymin=0,
    minor tick num=1,
    minor grid style={draw=gray!25},
  },
}

\pgfplotsset{
  DeletePlot/.style={
    small,
    width = 0.8\linewidth,
    height=0.50\linewidth,
    xlabel={Filter Size},
    ylabel={Throughput (M/s)},
    xlabel near ticks,
    x label style={at={(0.5,-0.12)},font=\small},
    scaled y ticks = base 10:-3,
    ylabel near ticks,
    y label style={at={(-0.1,0.5)},font=\small},
    ytick style={draw=none},
    ytick scale label code/.code = {},
    yticklabel style={ /pgf/number format/fixed },
    ymajorgrids,
    yminorgrids,
    minor tick num=1,
    minor grid style={draw=gray!25},
  },
}

\pgfplotsset{
  HistoPlot/.style={
    footnotesize,
    ybar = 1pt,
    bar width = 3.5pt,
    width = \linewidth + 30pt,
    height = 1.8in,
    ymin = 0,
    xmin = 100,
    xmax = 20000,
    xtick=data,
    xticklabels={100\,\si{\nano\second}, 250\,\si{\nano\second}, 500\,\si{\nano\second}, 1\,\si{\micro\second}, 2.5\,\si{\micro\second}, 5\,\si{\micro\second}, 10\,\si{\micro\second}, >10\,\si{\micro\second}},
    x tick label style = {rotate = 45, anchor = 15, xshift = 2pt},
    ytick scale label code/.code = {},
    yticklabel style={
        /pgf/number format/fixed,
    },
    scaled y ticks = base 10:-6,
    enlarge x limits = 0.08,
    fill,
    no marks,
    nodes near coords,
    every node near coord/.append style={rotate = 90, anchor = west, font=\tiny,/pgf/number format/fixed relative,/pgf/number format/precision=2},
    nodes near coords={
       \pgfkeys{/pgf/fpu}%
       \pgfmathparse{\pgfplotspointmeta / 640000}%
       \pgfmathprintnumber{\pgfmathresult}\%
    },
  },
}

\pgfplotsset{
  MHMLFPlot/.style={
    ybar,
    bar width = 5.5,
    no marks,
    width=\columnwidth,
    height=0.6\columnwidth,
    ylabel ={Load Factor},
    ymin = 0,
    ymax = 1,
    symbolic x coords = {wa_gqf_32, wa_gqf_64, wa_nogqf_64, r_gqf_14, r_gqf_24, r_nogqf_24},
    xticklabels = {{}, WA \\GQF N32, WA \\GQF N64, WA \\NoGQF N64, Rhizo \\GQF N14, Rhizo \\GQF N24, Rhizo \\NoGQF N24},
    x tick label style = {font = \small, align = center, rotate = 50, anchor = north east},
    major x tick style = transparent,
    enlarge x limits = 0.2,
    ymajorgrids,
    yminorgrids,
    minor y tick num = 1,
    fill,
    nodes near coords,
    every node near coord/.append style =
    {   
      rotate = 90, 
      anchor = west,
      fill = white,
      outer sep = 0,
      inner sep = 1,
      xshift = 1pt,
      /pgf/number format/fixed,
      /pgf/number format/precision = 1,
      font=\scriptsize
    }   
  },  
}

\pgfplotsset{
  MHMMemPlot/.style={
    ybar stacked,
    bar width = 12,
    no marks,
    width=\columnwidth,
    height=0.6\columnwidth,
    ylabel ={Min memory required (GB)},
    ymin = 0,
    ymax = 2500,
    symbolic x coords = {wa_gqf, wa_nogqf, r_gqf, r_nogqf},
    xticklabels = {{}, WA \\GQF, WA \\NoGQF, Rhizo \\GQF, Rhizo \\NoGQF},
    x tick label style = {font = \small, align = right, rotate = 0, anchor = north east},
    major x tick style = transparent,
    enlarge x limits = 0.2,
    ymajorgrids,
    yminorgrids,
    minor y tick num = 1,
    fill,
    nodes near coords, 
    every node near coord/.append style =
    { 
      anchor = west,
      fill = white,
      outer sep = 0,
      inner sep = 1,
      /pgf/number format/fixed,
      /pgf/number format/precision = 1,
      font=\small
    } 
  },  
}

\pgfplotsset{
  GQF_LFStyle/.style     = {color = teal,       mark = triangle*},
  HT_ALFStyle/.style     = {color = orange,       mark = triangle*},
  HT_MLFStyle/.style     = {color = red,       mark = triangle*},
}

\pgfplotsset{
  MHMMemPlot/.style={
    ybar stacked,
    bar width = 12,
    no marks,
    width=\columnwidth,
    height=0.6\columnwidth,
    ylabel ={Min memory required (GB)},
    ymin = 0,
    ymax = 2700,
    symbolic x coords = {wa_gqf, wa_nogqf, r_gqf, r_nogqf, tyme_gqf, tyme_nogqf},
    xticklabels = {{}, WA \\TCF, WA \\NoTCF, Rhizo \\TCF, Rhizo \\NoTCF, TYME \\TCF, TYME \\NoTCF},
    x tick label style = {font = \small, align = center, rotate = 0, anchor = north},
    major x tick style = transparent,
    enlarge x limits = 0.15,
    ymajorgrids,
    yminorgrids,
    minor y tick num = 1,
    fill,
    nodes near coords, 
    every node near coord/.append style =
    { 
      anchor = west,
      fill = white,
      outer sep = 0,
      inner sep = 1,
      /pgf/number format/fixed,
      /pgf/number format/precision = 1,
      font=\small
    } 
  },  
}

\colorlet{ColorLightTeal}{teal!60}

\pgfplotsset{
  QFNLStyle/.style     = {color = teal,       mark = square*},
  QFLStyle/.style     = {color = orange,       mark = triangle*},
  sortedStyle/.style     = {color = red,       mark = diamond*},
  constStyle/.style     = {color = violet,       mark = asterisk},
  dynamicStyle/.style     = {color = cyan,       mark = pentagon*},
  staticStyle/.style     = {color = gray,       mark = pentagon*},
  BloomStyle/.style     = {color = gray,       mark = triangle*},
  SQFStyle/.style     = {color = orange,       mark = pentagon*},
  RSQFStyle/.style     = {color = violet,       mark = diamond*},
  BBloomStyle/.style     = {color = magenta,       mark = diamond*},
  GQFStyle/.style     = {color = red,       mark = *},
  TCQFStyle/.style     = {color = teal,       mark = square*},
  88Style/.style     = {color = teal,       mark = square*},
  128Style/.style     = {color = orange,       mark = triangle*},
  1212Style/.style     = {color = red,       mark = diamond*},
  1216Style/.style     = {color = violet,       mark = asterisk},
  1632Style/.style     = {color = gray,       mark = triangle*},
  1232Style/.style     = {color = cyan,       mark = pentagon*},
  1616Style/.style     = {color = gray,       mark = pentagon*},
  1216UnalignedStyle/.style = {color = orange,       mark = pentagon*},
  MHMFIGStyle/.style = {color=ColorLightTeal, mark = triangle*}
}

\title{High-Performance Filters for GPUs}
%

%

\settopmatter{authorsperrow=4}

\author{Hunter McCoy}
\affiliation{%
  \institution{University of Utah}
  \country{USA}
}
\email{hunter@cs.utah.edu}

\author{Steven Hofmeyr}
\affiliation{%
  \institution{Lawrence Berkeley National Lab}
  \country{USA}
}
\email{shofmeyr@lbl.gov}

\author{Katherine Yelick}
\affiliation{%
  \institution{University of California Berkeley}
  \country{USA}
}
\email{yelick@berkeley.edu}

\author{Prashant Pandey}
\affiliation{%
  \institution{University of Utah}
  \country{USA}
}
\email{pandey@cs.utah.edu}


\begin{abstract}

Filters approximately store a set of items while trading off accuracy for space-efficiency and can address the limited memory on accelerators, such as GPUs. However, there is a lack of high-performance and feature-rich GPU filters as most advancements in filter research has focused on CPUs.

In this paper, we explore the design space of filters with a goal to develop massively parallel, high performance, and feature rich filters for GPUs. We evaluate various filter designs in terms of performance, usability, and supported features and identify two filter designs that offer the right trade off in terms of performance, features, and usability.

We present two new GPU-based filters, the TCF and GQF, that can be employed in various high performance data analytics applications. 
The TCF is a set membership filter and supports faster inserts and queries, whereas the GQF supports counting which comes at an additional performance cost.
Both the GQF and TCF provide point and bulk insertion API and are designed to exploit the massive parallelism in the GPU without sacrificing usability and necessary features.
The TCF and GQF are up to $4.4\times$ and $1.4\times$ faster than the previous GPU filters in our benchmarks and at the same time overcome the fundamental constraints in performance and usability in current GPU filters.

\end{abstract}


\maketitle

\section{Introduction}

As scientific and commercial data sets explode in volume and data rates, some of the high performance data analytics pipelines take advantage of the massive parallelism and advanced computing architecture of the GPUs. GPUs have proven to be effective accelerators for machine learning and simulation problems~\cite{tensorflow2015-whitepaper, gslide}, database engines~\cite{bress2013time,strohm2015gpu,wang2014concurrent,patta2015enhancing,li2016hippogriffdb,bress2014gpu}, and large-scale genomics pipelines~\cite{GeorganasEHG18,HofmeyrEGC20,sitaridi2016gpu,kobus2021metacache,besta2020communication}.


In this paper, we consider the use of GPUs in filtering, one of the key operations in many data processing and analytics pipelines. GPUs offer both an opportunity for performance improvement and a challenge for data analytics due to the limited GPU memory that is available.

\emph{Filters}, such as Bloom~\cite{Bloom70}, quotient~\cite{PaghPaRa05,DillingerMa09,EinzigerFr16,BenderFaJo12a,PandeyBJP17,pandeySigmod21} and cuckoo filters~\cite{FanAnKa14,BreslowJ18}, maintain an approximate representation of a set or a multiset\footnote{Counting filters maintain count estimates of items in a multiset. A counting filter may have an error rate $\delta$. Queries return true counts with probability at least $1 - \delta$. Whenever a query returns an incorrect count, it must always be greater than the true count. Counting filters offer no guarantee on the overestimate unlike count sketches. We refer the readers to Goswami et al's~\cite{goswami2018buffered} paper for a detailed comparison.}. The approximate representation saves space by allowing queries to occasionally return a false-positive. For a given false-positive rate $\epsilon$: a membership query to a filter for set $S$ returns present for any $x \in S$, and returns absent with probability at least $1 - \epsilon$ for any $x \notin S$. A filter for a set of size $n$ uses space that depends on $\epsilon$ and $n$ but is much smaller than explicitly storing all items of $S$.

Given the popularity and wide-scale impact of filters there have been many papers in the last decade that advance the theory and practice of filters~\cite{BenderFaJo12a,PandeyBJP17,pandeySigmod21,FanAnKa14,BreslowJ18,QiaoLiCh14,putze2007cache,LuDeDu11,CanimMiBh10,DebnathSeLi11, AlmeidaBaPr07,FanCaAl00,BonomiMiPa06,CostaAR09,IacobIS15,PaghPaRa05,DillingerMa09,EinzigerFr16}. Most of these papers have focused on improving the state of the art in terms of space usage and performance. A few papers have also explored adding new features in the filter such as deletion, associating small values with hashes, and counting, which are critical in many applications~\cite{BenderFaJo12a,PandeyBJP17,pandeySigmod21,FanAnKa14}.

However, there is very little work on building fast, space-efficient, and feature-rich filters for the GPU. Costa et al.~\cite{CostaAR09} and Iacob et al.~\cite{IacobIS15} showed how to build and query Bloom filters on the GPU. Geil et al.~\cite{GeilFO18} first showed how to build and query a quotient filter on the GPU using the bulk build API. These filter implementations offer sub-optimal performance, do not offer choices in terms of space usage and false-positive rate trade-off, and do not have adequate APIs to be integrated in many data analytics applications. Furthermore, these implementations do not support critical features such as deletion, counting, or associating values with the items, which are required by modern-day applications~\cite{PandeyBJP17a,PandeyBJP17b,PandeyABFJP18Cell}. Due to the lack of available options modern GPU-accelerated applications often work around the limitations of filters which in turn results in sub-optimal use of resources and further hinders their scalability to larger datasets.

For example, MetaHipMer~\cite{GeorganasEHG18,HofmeyrEGC20} is an extreme-scale \emph{de novo} metagenome assembler that leverages GPUs to speed up raw data processing and is designed to scale out to thousands of nodes to handle terabyte scale data. MetaHipMer requires a filter that can map fingerprints to small values to weed out singletons during raw data processing and use the output in later stages of the pipeline. It cannot use Bloom filters since Bloom filters do not support associating small values with the items. Similar to the \bloom, Geil et al.'s~\cite{GeilFO18} quotient filter (SQF) cannot associate values. In addition, the SQF can only scale up to a few million items and does not offer the right trade off in terms of space usage and false-positive rate. Similarly, many database engines~\cite{krueger2011applicability,kozawa2012gpu,strohm2015gpu} that leverage GPUs to speed up merge and join operations cannot use existing filters as they do not support counting and enumeration of items which are required for merge and join operations.

Designing filters for GPUs comes with a host of unique challenges. The architecture of GPUs, originally designed to accelerate rendering operations, provides massive parallelism at the expense of limited memory, simpler instructions, and synchronization tools.
The differences in the architecture between CPUs and GPUs cause filters designed for CPUs to often have sub-optimal performance when directly ported to GPUs. Thread contention and thrashing are often issues for GPU data structures, even those with large sizes. In addition, the grouping of threads into warps is another constraint on the design, with grouped access patterns providing massive speed boosts over naive implementations. The random access pattern of many existing filters amplifies this problem as threads in a warp are likely to diverge while accessing memory locations randomly throughout the filter.




\para{Our contribution}
In this paper, we explore the design space of filters and identify the designs that can exploit the massive parallelism on the GPU without introducing fundamental feature limitations and giving up performance and usability. We identify two filter designs that offer appropriate trade-offs in terms of the performance and necessary features. Based on these designs, we develop and evaluate two new GPU filters, the two-choice filter (TCF) and the GPU-based counting quotient filter (GQF). The TCF does not support counting which enables faster inserts and queries, whereas the GQF supports counting at an additional performance cost. Both filters support deletions and associating small values with fingerprints.


\ppoppaddition{
The TCF is designed to organize fingerprints in blocks sized to fit inside a GPU cache line. It uses Cuda cooperative groups to perform insert, query, and delete operations inside these blocks to achieve massive parallelism without any contention. The TCF further uses the power-of-two-choice (POTC) hashing~\cite{AzarSTOC94} to minimize the load imbalance across fixed-size blocks and achieve a high load factor. However, the block sizes in TCF are smaller than what is used in the CPU filter~\cite{pandeySigmod21}. This leads to increased load variance across blocks resulting in lower maximum achievable load factor. To overcome this problem, the TCF uses a backing store to achieve high load factors. To the best of our knowledge, the TCF is the first filter design that uses a backing store.
}

\ppoppaddition{
The TCF combines the POTC hashing, backing store, and cooperative groups in just the right way and shows a new algorithmic paradigm to achieve a stable filter design.
}

The TCF strips out the ability to count in favor of faster inserts and query operations. The TCF offers both concurrent inserts and queries and bulk insertion API. The TCF can represent a set of items approximately and supports deletions, enumeration, and associating small values with items.



The GQF is a GPU-optimized implementation of the counting quotient filter~\cite{PandeyBJP17a}.  The GQF is designed to overcome the fundamental limitations of the earlier implementation of the quotient filter~\cite{GeilFO18} on the GPU, such as only supporting a fixed false-positive rate and scaling only to a few million items. The GQF offers all the features that modern data analytics applications demand, e.g., better space-accuracy trade-off, counting, deletions, associating values with items, and resizability. In addition, it offers both concurrent inserts and queries and a bulk insert API, unlike the earlier GPU-based filter implementations.

\ppoppaddition{
The GQF demonstrates a novel coordinated lock-free implementation for batch insertions. The lock-free implementation partitions the filter into exclusive-access even-odd regions and assigns threads inside a warp to fixed memory regions to achieve low thread divergence and avoid thrashing. 
We believe that our even-odd scheme for bulk insertions can also be applied to other linear-probing-based hash tables to accelerate insertions~\cite{pandey2022iceberght} and also for storing dynamic graphs on GPUs~\cite{pandey2021terrace,pandey2021variantstore}.
}

\para{Our results}
The TCF and GQF offer far better (up to three orders of magnitude in some cases) performance and use less or similar space than other filters on the GPU offering a smaller set of features.
\begin{enumerate}[noitemsep, leftmargin=*]
    \item The point TCF  is up to $4.45\times$ faster for inserts and queries than all filters that support deletions.
    \item The GQF is up to $1.93\times$ and $2.4\times$ faster than the GPU-based Bloom filter for inserts and queries respectively.
    \item The bulk TCF achieves an insertion throughput of $3.4$ Billion items per second on NVIDIA A100 GPUs.
    \item The bulk TCF achieves an insertion throughput of 70\% of the Blocked Bloom filter with half the false positive rate.
    \item The TCF is over an order of magnitude faster than all other filters for deletions.
    \item The GQF supports high throughput counting (800+ Million/sec) on both simulated and real-world datasets.
\end{enumerate}

\section{A Brief History of Filters}\label{sec:prelim}

%
%

In this paper, we consider dynamic filters as they have wide-spread applications in data analytics.
Dynamic filters approximately represent a set of items that does not need to be known before the construction. Dynamic filters have seen much more advancement in the last few decades as applications often do not know the set of items in advance. Examples of dynamic filters are Bloom filters~\cite{Bloom70}, quotient filters~\cite{BenderFaJo12, PandeyBJP17b,DillingerMa09,PaghPaRa05,EinzigerFr16}, and cuckoo filters~\cite{FanAnKa14,BreslowJ18}.

\textbf{Bloom filters} consume $\log(e)\, \opt$ space, which is roughly $\log(e)\approx 1.44$ times more than the lower bound of $\opt + \Omega(n)$ bits~\cite{CarterFG78}. In contrast, for a set $S$ taken from a universe $U$, where $|U|=u$, an error-free dictionary requires $\Omega(\log {u\choose n}) \approx \Omega(n \log u)$ bits. Bloom filters also incur $\errbits$ cache-line misses on inserts and positive queries, giving them poor insertion and query performance.

\textbf{Blocked Bloom filters}~\cite{putze2007cache} overcome the poor cache locality of Bloom filters by constructing a series of smaller Bloom filters each of which is small enough to fit inside a small number of cache lines. The first hash function is used to select a block and rest of the hash functions are used to set/test bits inside the block. However, the cache efficiency comes at the cost of higher false-positive rate. Blocked Bloom filters have theoretically and empirically higher (up to $5\times$) false positive rates compared to Bloom filters. See \Cref{tab:merged_fp_space} for the empirical calculations of FP rate.

\textbf{Quotient filters}~\cite{Cleary84,PaghPaRa05,DillingerMa09,BenderFaJo12a,PandeyBJP17,pandeySigmod21} represent a set approximately by compactly storing small fingerprints of the items in the set via Robin Hood hashing~\cite{CelisLaMu85}. The quotient filter uses $1.053 (2.125 + \log_21/\epsilon)$ bits per element, which is less than the Bloom filter whenever $\epsilon \leq 1/64$, which is the case in almost all applications. It supports insertion, deletion, lookups, resizing, and merging. The counting quotient filter (CQF)~\cite{PandeyBJP17}, improves upon the performance of the quotient filter and adds variable-sized counters to count items using asymptotically optimal space, even in large and skewed datasets. In the counting quotient filter, we can also associate small values with items either by re-purposing the variable-sized counters~\cite{PandeyABFJP18Cell} to store values or by explicitly storing small values with the remainders in the table~\cite{PandeySMB20}.

\textbf{Cuckoo filters}~\cite{FanAnKa14,BreslowJ18} also store small fingerprints compactly in a table. However, unlike the quotient filter that uses Robin Hood hashing, the cuckoo filter uses cuckoo hashing to resolve collisions among fingerprints. Cuckoo hashing uses kicking (or cuckooing) to find an empty slot for the new item when all the slots in a bucket are occupied. This results in a cascading sequence of kicks until the filter converges on a new stable state. Inserts become slower as the structure becomes full, and in fact inserts may fail if the number of kicks during a single insert exceeds a specified threshold (500 in the author's reference implementation). 

\textbf{Two-Choice filters}~\cite{pandeySigmod21} organize fingerprints compactly in blocks similar to the cuckoo filter. However, unlike the cuckoo filter, there is no kicking. The blocks in the two-choice filter are larger in size ($\approx \log{n}$, where $n$ is the number of items which is usually the size of the cache line on most machines) than the cuckoo filter and power-of-two-choice hashing is used to reduce the variance across the blocks and achieve a high load factor. During insertions if both blocks corresponding to a fingerprint are full then the data structure is declared full. The power-of-two-choice hashing enables the filter to probe exactly two cache lines during inserts and queries and write to a single cache line during inserts. Given the larger block sizes the vector quotient filter~\cite{pandeySigmod21} uses quotienting (similar to the quotient filter) to organize fingerprints inside blocks. It divides the fingerprints into a quotient and remainder part and only stores the remainder in the slot given by the quotient. It uses two additional metadata bits to resolve collisions among quotients.



\section{Designing a GPU filter}\label{sec:design}

Here we discuss the design principles needed to build a fast and space efficient filter on the GPU and use them to analyze various filter designs.


\subsection{GPU design principles}
There are four major design principles to consider when implementing data structures on GPUs:

\begin{enumerate}[noitemsep, leftmargin=*]
    \item \textbf{Low thread divergence:} threads inside a warp should execute the same instruction. This enables writing simple kernels that can exploit massive parallelism in the GPU.
    \item \textbf{High memory coherence:} threads inside a warp should access the same  memory from a local region. Random memory accesses are expensive and cause threads to stall.
    \item \textbf{High degree of parallelism:} a high number of threads saturate memory bandwidth and hide memory latency. 
    \item \textbf{Atomic operations:} atomic operations help efficient thread scheduling inside a warp. Non-atomic writes and data movements cause slow downs and require locking large memory regions. Locking results in high overheads and affects the overall throughput.
\end{enumerate}



\subsection{Analysis of filter designs}

We now look at the dynamic filters discussed in~\Cref{sec:prelim} and evaluate them based on the GPU design principles. 

Bloom filters are easy to implement on the GPU as they only require test and set operations. These operations can be implemented using atomic operations and achieve low thread divergence. However, each operation results in multiple cache misses and therefore Bloom filters have low memory coherence. They also have sub-optimal space usage. Moreover, Bloom filters do not support deletions, counting\footnote{The counting Bloom filter~\cite{FanCaAl00}, a variant of the Bloom filter, supports counting but it comes at a high space-overhead which makes it highly inefficient in practice.}, and associating small values with items.

Blocked Bloom filters are better suited to GPUs.
Each operation requires probing inside a single block. They achieve low thread divergence, high memory coherence, a high degree of parallelism, and atomic operations. Thus, blocked Bloom filters can satisfy all the GPU design principles. However, blocked Bloom filters have a high false-positive rate compared to Bloom filters and also do not support necessary features like deletions and counting.

Operations in the quotient filter have high cache locality which makes it an appropriate choice to achieve high memory coherence. However, insert operations in the quotient filter requires shifting fingerprints which makes it harder to use atomic operations and also results in high thread divergence. However, the quotient filter can support all the necessary features like deletions, counting, and associating small values with items which makes the quotient filter a highly usable data structure that multiple applications can benefit from. 

It is quite challenging to achieve high speed operations while maintaining all of the features in a GPU implementation of the quotient filter. Geil et al~\cite{GeilFO18} implemented a preliminary version of the GPU quotient filter. However, that implementation was adapted from Bender et al.'s quotient filter~\cite{BenderFaJo12a}, which did not have all the features, like counting and value association, and also had higher space overhead. Furthermore, Geil et al's GPU-based quotient filter has implementation-specific limitations (e.g., it supports a fixed false-positive rate and can only be sized to store less than $2^{26}$ items) resulting in poor performance and limited scalability.

The cuckoo filter stores fingerprints in fixed size blocks. This design is amenable to high memory coherence and low thread divergence. Atomic operations can also be used to read and write fingerprints. However, the cascading sequence of reads and writes to random memory locations makes the cuckoo filter hard to implement efficiently on the GPU. In particular, at high load factors when the number of kicked items becomes high, each insertion will result in very low memory coherence. Moreover, each kicking operation results in multiple cache-line writes. This makes it challenging to achieve high speed operations in a GPU cuckoo filter. Moreover, cuckoo filters do not support counting and associating small values with items.

The two-choice filter has the advantages of the cuckoo filter design. It has fixed size blocks. Each operation requires probing into exactly two blocks, and inserts and deletes only write into a single block. This results in low thread divergence, high memory coherence, and a high degree of parallelism. However, due to large block sizes a more sophisticated structure is required to maintain fingerprints inside each block. Therefore, it is not straightforward to use atomic operations to read or write fingerprints inside blocks. It is a challenging task to implement a two choice filter on the GPU using atomic operations to achieve high throughput.

\subsection{Most efficient GPU filter designs}

We now identify the filters that offer necessary features and can achieve high speed operations on the GPU. First, we pick the two-choice filter (TCF). The TCF achieves three out of four design principles. It achieves low thread divergence, high memory coherence, and high degree of parallelism. It also supports deletions unlike the Bloom filter variants. We redesign the TCF to use atomic operations and cooperative groups to exploit massive GPU parallelism. Second, we pick the counting quotient filter (CQF). The CQF offers all the necessary features that modern applications demand. In particular, it supports counting and value associations which are critical features for many applications. However, it is hard to achieve low thread divergence and high parallelism in the CQF. We will redesign the CQF to use a coordinated lock-free approach and achieve massive parallelism and scalability.


\section{TCF Implementation}\label{sec:TCF_impl}

\begin{figure}
    \centering
    \resizebox{\columnwidth}{!}{%
    \includegraphics{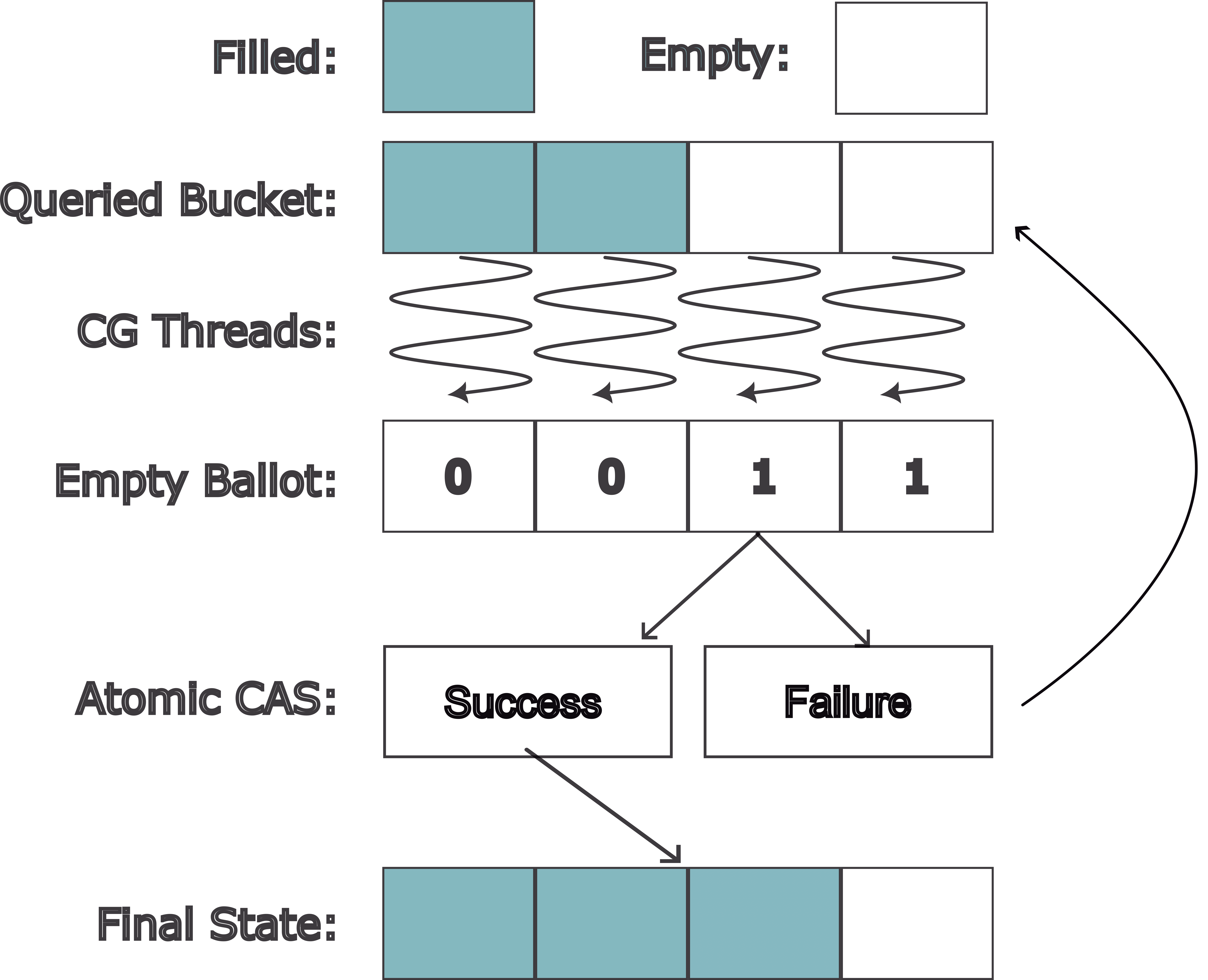}
    }
    \caption{Block insertion in the TCF using cooperative groups.}
    \label{fig:tcf_fig}
\end{figure}

In this section, we give the implementation details of the two choice filter (TCF) on the GPU. We first explain the version that supports concurrent inserts and queries via the use of atomics. We then explain the bulk lock-free version that utilizes sorting to precondition items for faster operations.

\ppoppaddition{
In the TCF, we organize the table into blocks. Each block can store $B$ $f$-bit fingerprints. The blocks are sized to fit inside a GPU cache line. The TCF uses the power-of-two-choice (POTC) hashing scheme to perform operations.
In a POTC scheme, every item is assigned two blocks via a pair of unique hashes. For inserts, the fill of each block is queried, and the item is inserted into the less full block. Queries return true if the queried item is found in either block. 
The POTC hashing helps to reduce the load variance across blocks, reducing the size of the largest block to $O(\log \log n)$, where $n$ is the number of items, as shown by Azar et al.~\cite{AzarSTOC94}.
}

Inserts and queries inside a block are performed using cooperative groups. A group cooperatively loads the block into shared memory before striding over the block to check for empty slots or the presence of an item. Once an empty slot has been found, the cooperative group ballots for a leader who will attempt an atomicCAS operation to write the item to global memory. On success, the cooperative group returns, while on failure the group will look for a new empty slot within the block and re-ballot to determine the new leader.
\ppoppaddition{Please refer to \Cref{alg:tcfBucket} and \Cref{fig:tcf_fig} for more details on block operations.}

\begin{algorithm}[t]
	\caption{Block\_Insert(Key, Val, CG)} \label{alg:tcfBucket}
    \begin{algorithmic}[1]

    \For{\texttt{i = CG.thread\_rank(); i < bucket\_len; i+=CG.size()}}

        \State bool ballot = 0;
        \If {\texttt{bucket[i] == empty OR bucket[i] == tombstone}}
            \State ballot = 1;

        \EndIf

        \State ballot\_result = CG.ballot(ballot);

        \While \texttt{ballot != 0}

            \If {CG.thread\_rank() == \_\_ffs(ballot) -1}

                \If  {\texttt{atomicCAS(bucket+i, EMPTY, Key - Val)}}

                \State CG.ballot(true)
                \State \Return true

                \Else

                \State CG.ballot(false)

                \EndIf
                
            \Else

                \If {CG.ballot(false)} \Comment{If current leader inserts, return}

                \State \Return true

                \EndIf
            \EndIf

            \State ballot = ballot $\oplus$ $1 << $ \_\_ffs(ballot) -1

        \EndWhile

        \EndFor

        \State \Return false \Comment{No slots were available.}
	\end{algorithmic} 
\end{algorithm}

\subsection{TCF design optimization}

Three factors that dominate the TCF performance: size of the blocks, the bits per item, and size of the cooperative groups.


The size of the blocks determines the number of cache line access during operations. Therefore, we enforce that the size of a block $\leq 128$ bytes (a cache line on GPU) which limits the number of accesses to two for the majority of operations.

The false-positive rate for the TCF is given by $\frac{2 B}{2^{f}}$, where $B$ is the size of the blocks and $f$ is fingerprint size. A larger fingerprint size decreases the false-positive rate but increases the space. The minimum size for an atomicCAS transaction is 2 bytes. With keys set to the minimum CAS size and a block size of 16, the error rate is $.04\%$. However, most practical applications require the error rate to be around $0.1\%$. To achieve that error rate, we can either increase the block size or decrease the fingerprint size. Increasing the block size has a negative effect on performance as each thread needs to look at more data. 
Storing 12-bit fingerprints brings down the space usage but $50\%$ of inserts now require two atomic operations, and fingerprints can no longer fully occupy an atomic transaction, meaning that an atomicCAS could fail due to a change in bits outside of the slot being operated on. 

The size of the cooperative groups is particularly important to the performance of the filter design, as it provides a trade off between computational and memory efficiency inside of a warp.
Increasing the number of cooperative groups in a warp increases the number of cache lines that can be simultaneously scheduled for loading, but decreases the number of workers available per block. 
A more detailed analysis of this phenomenon, along with experimental results of varying the cooperative group size, are found in~\Cref{sec:tcf-variations}. 
%

\paragraph{Backing table}
To avoid insertion failures (no empty slot in both blocks) before reaching a 90\% load factor we use a backing table. We use a small double-hashing-based backing table sized to 1/100th of the size of the main table for storing any items that fail to be inserted. Since $<< 1\%$ items fail to be inserted, the extra cost required to insert and query from this table is negligible, and it has no measured effect on the speed of inserts or positive queries. However, it does have an effect on the performance of false-positive queries, as at least one extra block will have to be searched. The TCF can achieve 90\% load factor using the backing table.

\paragraph{Shortcut optimization}
As shown in Pandey et al. ~\cite{pandeySigmod21}, in the case where the primary block has a very low fill ratio, we can safely insert into the primary block without querying the secondary block. This reduces the number of cache loads required to insert by one, improving speed. After empirical testing, we found a $0.75$ fill ratio to be the ideal cutoff for this shortcut optimization, as it provided the best performance without affecting the variance between blocks.

\subsection{Bulk TCF}

The bulk version of the TCF utilizes sorting to increase the efficiency of read/write operations in the GPU. Like reads, writes on a GPU can be coalesced, with up to 128 bytes of contiguous memory being written in one operation.
For example, if all 16-bit keys that are going to be inserted into the same cache line can be aggregated before insertion, we could see up to a $32\times$ performance increase for memory operations. 

If the time saved on insertion is less than the cost of aggregating items, we can improve the throughput via an aggregation phase.
Items are sorted and passed to the bulk TCF as a sorted list of items to be inserted into a block.
Blocks of the TCF are loaded into shared memory before items are inserted and all reads and writes are performed using shared memory atomics. At the end, kernel writes occur as coalesced writes to global.
This minimizes the data written to global as all writes to global occur as cooperative cache-wide coalesced writes.

Unlike the point TCF, blocks in the bulk version maintain a sorted list of items inside the block. This allows the blocks to be queried in logarithmic time via a binary search, or in linear time for a batch of queries. To efficiently insert while maintaining a sorted order, each cooperative group maintains three lists during insertion: the list of items currently stored in the block, the sorted list of items that can be shortcutted into the block, and the list of items assigned to the block via POTC hashing.
The three lists are merged together using a parallel zip strategy, and the resulting block is cooperatively written to global memory.

The bulk filter has an error rate of $0.3\%$ with a block size of 128 and a 16 bits per item. While this is appropriate for most applications, it requires $33\%$ more space per item to achieve the same error rate as the point filter.

\section{GQF Implementation}\label{sec:gqf_impl}



In this section, we give an overview of Pandey et al.'s~\cite{PandeyBJP17a} counting \qf (CQF). We also describe the locking mechanism in the counting \qf for thread-safe operation because it acts as the building block in the \sysname (GQF). We finally explain how we design the counting quotient filter for the GPU.

\subsection{CQF overview}
The counting quotient filter (CQF) stores an approximation of a multiset $S\subseteq \mathcal{U}$ by storing a compact, lossless representation of the multiset $h(S)$, where $h:\mathcal{U}\rightarrow\{0,\ldots,2^p-1\}$ is a hash function that maps items from the universe $\mathcal{U}$ to a $p$-bit fingerprint. To handle a multiset of up to $n$ distinct items while maintaining a false-positive rate of at most $\epsilon$, the CQF sets $p=\log_2 \frac{n}{\epsilon}$ (see the original quotient filter paper for the analysis~\cite{BenderFaJo12a}).

\ppoppaddition{
The \cqf divides $h(x)$ into its first $q$ bits, \defn{quotient}
$h_0(x)$, and its remaining $r$ bits, \defn{remainder} $h_1(x)$.  It maintains an array $Q$ of $2^q$ $r$-bit slots, each of which can
hold a single remainder.  When an element $x$ is inserted, the \cqf attempts to store the remainder $h_1(x)$ at index $h_0(x)$ in $Q$
(which we call $x$'s \defn{canonical slot}).  If that slot is already in use, then the \cqf uses Robin hood hashing~\cite{PandeyBJ17} to find the next available empty slot to store $h_1(x)$. All the items that share the same canonical slot are stored together in a \defn{run} and a sequence of runs stored contiguously with no empty space is called a \defn{cluster}. During an insert operation, the next available empty slot is found at the end of the cluster. If an item lands at the start of the cluster then all the items in the cluster must be shifted to create an empty space.
}

\subsection{Point insertion API}

In the point implementation, each thread acquires exclusive access to a section of memory for writing.
Internal remainder shifts are processed using a custom {\it memmove} function, as the driver API only supports {\it memcpy}, which does not guarantee write safety when the source and destination regions overlap.

To perform an insert operation, the thread needs to lock a big enough region so that shifting items will not corrupt the subsequent region where another thread might be operating. Therefore, the slots are divided into locking regions that are big enough to handle the shifting of remainders during insertions without causing an overflow to the next locking region.
Given that the filter is only filled to 95\% load factor, we can safely say with that the maximum cluster size will be less than 8192 slots~\cite{PandeyBJP17a}. In order to guarantee that each insert has at least this many slots to work with, we divide the filter into sections of 8192 slots. An insert thread grabs two locks corresponding to the canonical slot of the item and the lock immediately after it. Locking two consecutive regions in the \cqf ensures that memory corruption bugs are avoided, even if we overflow into the next region during an insert operation. The insert thread holds these locks until all changes are flushed to memory.


The length of the longest cluster is bounded by $O(\frac{\ln{2^q}}{\alpha-\ln{\alpha}-1})$ with high probability~\cite{BenderFaJo12a, PandeyBJP17}, where $q$ is the number of quotient bits, $2^q$ is the number slots in the QF, and $\alpha$ is the load factor.
For example, if $q=40$ (i.e., $2^{40}$ slots) and $\alpha =3/4$, the largest cluster in the filter has 736 slots. On average, clusters are $O(1)$ in size.


To perform an operation, CUDA atomics require exclusive access to a cache line's worth of memory, e.g., 128 bytes on the Tesla V100s. With one bit per lock, there would be 1024 locks in a cache line.
This would lead to heavy thread contention 
and cause the vast majority of threads to thrash. To ameliorate this, we used cache-aligned locks, as the number of locks relative to the total size of the data structure is small enough that they only contribute a small percentage to the overall space usage.

\begin{figure}
    \centering
    \resizebox{\columnwidth}{!}{%
    \includegraphics{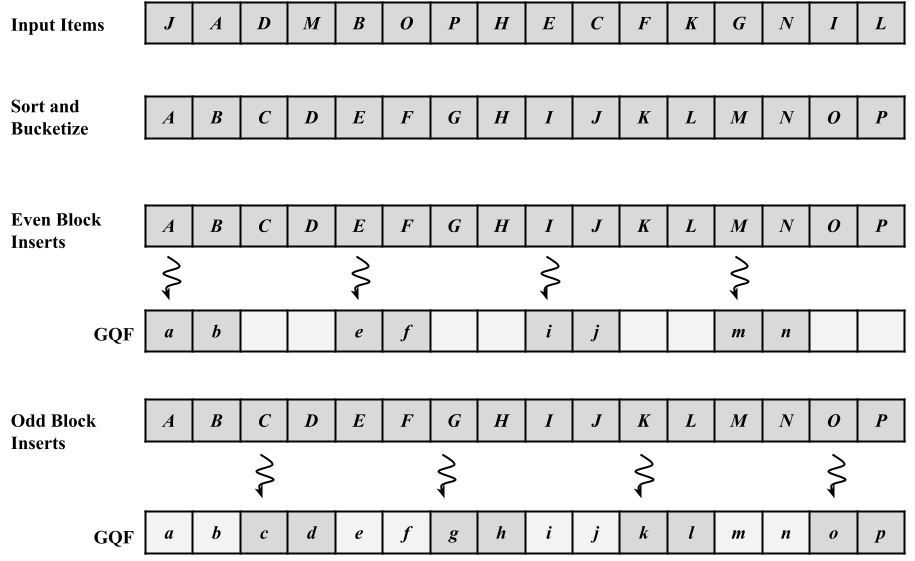}
    }
    \caption{Bulk inserts in two phases in the GQF.}
    \label{fig:gqf_inserts}
\end{figure}

\subsection{Bulk insertion API}


In the bulk API, we group items that hash to the same region and a single thread is assigned to each region for inserting all the grouped items. This guarantees that threads will have exclusive access to regions. 

To avoid the overhead of locking, we perform the insert operation in two phases. In the first phase, items belonging to even regions are inserted, with each thread assigned a specific region. Since there are no threads operating in the odd regions we can safely perform insertions without any memory corruption issues. In the second phase, the items belonging to the odd regions are inserted. 

This "Even-odd region" scheme maximizes the number of inserts that can safely occur simultaneously. Although it only allows insertion into half of the regions during a given phase, for large filter sizes the number of regions far exceeds the number of threads, allowing for full saturation of the GPU. Each region is sized to 8192 slots and phased insertion guarantees that threads are $\approx16$K slots apart and will find empty slots before overflowing into the next region.
\ppoppaddition{Please refer to~\Cref{fig:gqf_inserts}.}


Our implementation of this insert scheme uses temporary \defn{buffers} to hold items corresponding to each region. To efficiently distribute items into regions we use atomic operations to set the buffer sizes and assign each item an index in the buffer. In practice, we do not allocate temporary buffers. Instead, we use pointers into the input array to mark the boundaries for the buffers. This saves memory and the time required to allocate memory at run time.

\paragraph{Sorting hashes}
Internally, the GQF stores items akin to a linear hash table, with the remainders in a run in sorted order. New items inserted into a run must therefore shift any remainders greater than them in order to maintain the sorted structure. These shifts are the dominate the insertion time. 
We can avoid these memory shifts by inserting remainders (or hashes) in a sorted order. If the entire dataset were sorted before insertion, no shifts would be required as each new remainder will be stored in the last empty slot.
A variant of this holds true when the input dataset is batched: while it is impossible to avoid shifting items already in memory, sorting the input batch removes any extraneous memory shifts of items in the current batch.



Our implementation uses the Thrust library \cite{thrustMisc} to perform an in-place sort on the input data. After sorting, the starts of buffers are set using successor search which finds the index of the smallest item greater than or equal to the minimum hash of the current buffer. This eliminates the need to use atomics to set the buffers which in turn saves time during multiple phases of insertion. 



\subsection{Optimization for skewed distributions}
\label{sec:skew}

Datasets with skewed distributions (where counts of the items are derived from a power-law or a Zipfian distribution \cite{Corominas_Murtra_2010}) cause high contention among threads in the point insert API and load imbalance in the bulk insert API. This results in much slower insertion throughput and limited scaling with increasing filter sizes.



For the bulk insert API, we take the map-reduce approach to avoid the high contention. We first sort the batch of input items and then perform a reduction to compress the duplicate items into $\langle\text{item},\; \text{count}\rangle$ pairs. This reduction allows us to perform a single insertion with the aggregate count for every item in the batch instead of multiple insertions corresponding to each instance of repeated item. This amortizes the cost of acquiring locks and performing insertions. It further enables us to reduce the load imbalance across regions resulting in high insertion throughput. In our implementation, mapping and reduction are handled by the Thrust library~\cite{thrustMisc}.

\section{Evaluation}\label{sec:eval}

\begin{figure*}
   \centering
   \ref{PointAgg}\\
   \begin{subfigure}{0.32\linewidth}
      \begin{tikzpicture}
         \begin{axis}[
               AggregatePlotClipped,
               legend entries={TCF, GQF, Bloom, Blocked Bloom},
               legend columns=6,
               legend to name={PointAgg}
            ]
            \addplot[TCQFStyle] table[x=x_0, y=y_0] {data/tcqf/tcqf_agg_insert.txt};
            \addplot[GQFStyle] table[x=x_0, y=y_0] {data/point_gqf/point_agg_insert.txt};
            \addplot[BloomStyle] table[x=x_0, y=y_0] {data/bloom/bloom_agg_insert.txt};
            \addplot[BBloomStyle] table[x=x_0, y=y_0] {data/blocked_bloom/blocked_bloom_agg_insert.txt};
           
         \end{axis}
      \end{tikzpicture}
      \caption{Cori Point Inserts.}
      \label{point-inserts}
   \end{subfigure}%
   \begin{subfigure}{0.32\linewidth}
      \begin{tikzpicture}
         \begin{axis}[
               AggregatePlotClipped
            ]
            \addplot[TCQFStyle] table[x=x_0, y=y_0] {data/tcqf/tcqf_agg_lookup.txt};
            \addplot[GQFStyle] table[x=x_0, y=y_0] {data/point_gqf/point_agg_lookup.txt};
            \addplot[BloomStyle] table[x=x_0, y=y_0] {data/bloom/bloom_agg_lookup.txt};
            \addplot[BBloomStyle] table[x=x_0, y=y_0] {data/blocked_bloom/blocked_bloom_agg_lookup.txt};
         \end{axis}
      \end{tikzpicture}
      \caption{Cori Point Positive Queries.}
      \label{point-exists}
   \end{subfigure}%
   \begin{subfigure}{0.32\linewidth}
      \begin{tikzpicture}
         \begin{axis}[
              AggregatePlotClipped
            ]
            
            \addplot[TCQFStyle] table[x=x_0, y=y_0] {data/tcqf/tcqf_agg_fp.txt};
            \addplot[GQFStyle] table[x=x_0, y=y_0] {data/point_gqf/point_agg_fp.txt};
            \addplot[BloomStyle] table[x=x_0, y=y_0] {data/bloom/bloom_agg_fp.txt};
            \addplot[BBloomStyle] table[x=x_0, y=y_0] {data/blocked_bloom/blocked_bloom_agg_fp.txt};
            
         \end{axis}
      \end{tikzpicture}
      \caption{Cori Point Random Queries.}
      \label{point-false}
   \end{subfigure}%
   \\
   \begin{subfigure}{0.32\linewidth}
      \begin{tikzpicture}
         \begin{axis}[
            AggregatePlotClippedPerl
         ]
            \addplot[TCQFStyle] table[x=x_0, y=y_0] {perlmutter_data/tcqf/tcqf_agg_inserts.txt};
            \addplot[GQFStyle] table[x=x_0, y=y_0] {perlmutter_data/point/insert.txt};
            \addplot[BloomStyle] table[x=x_0, y=y_0] {perlmutter_data/bloom/bloom-aggregate-insert.txt};
            \addplot[BBloomStyle] table[x=x_0, y=y_0] {perlmutter_data/blocked_bloom/bloom_agg_insert.txt};
           
         \end{axis}
      \end{tikzpicture}
      \caption{Perlmutter Point Inserts.}
      \label{perl-point-inserts}
   \end{subfigure}%
   \begin{subfigure}{0.32\linewidth}
      \begin{tikzpicture}
         \begin{axis}[
               AggregatePlotClippedPerl
            ]
            \addplot[TCQFStyle] table[x=x_0, y=y_0] {perlmutter_data/tcqf/tcqf_agg_lookups.txt};
            \addplot[GQFStyle] table[x=x_0, y=y_0] {perlmutter_data/point/lookup.txt};
            \addplot[BloomStyle] table[x=x_0, y=y_0] {perlmutter_data/bloom/bloom-aggregate-exists-lookup.txt};
            \addplot[BBloomStyle] table[x=x_0, y=y_0] {perlmutter_data/blocked_bloom/bloom_agg_lookup.txt};
         \end{axis}
      \end{tikzpicture}
      \caption{Perlmutter Point Positive Queries.}
      \label{perl-point-exists}
   \end{subfigure}%
   \begin{subfigure}{0.32\linewidth}
      \begin{tikzpicture}
         \begin{axis}[
              AggregatePlotClippedPerl
            ]
            
            \addplot[TCQFStyle] table[x=x_0, y=y_0] {perlmutter_data/tcqf/tcqf_agg_fp.txt};
            \addplot[GQFStyle] table[x=x_0, y=y_0] {perlmutter_data/point/fp.txt};
            \addplot[BloomStyle] table[x=x_0, y=y_0] {perlmutter_data/bloom/bloom-aggregate-false-lookup.txt};
            \addplot[BBloomStyle] table[x=x_0, y=y_0] {perlmutter_data/blocked_bloom/bloom_agg_fp.txt};
            
         \end{axis}
      \end{tikzpicture}
      \caption{Perlmutter Point Random Queries.}
      \label{perl-point-false}
   \end{subfigure}%
   \vspace{-0.5em}
\caption{Point API aggregate throughput comparison between various filters.}
\label{fig:point-performance}
\end{figure*}
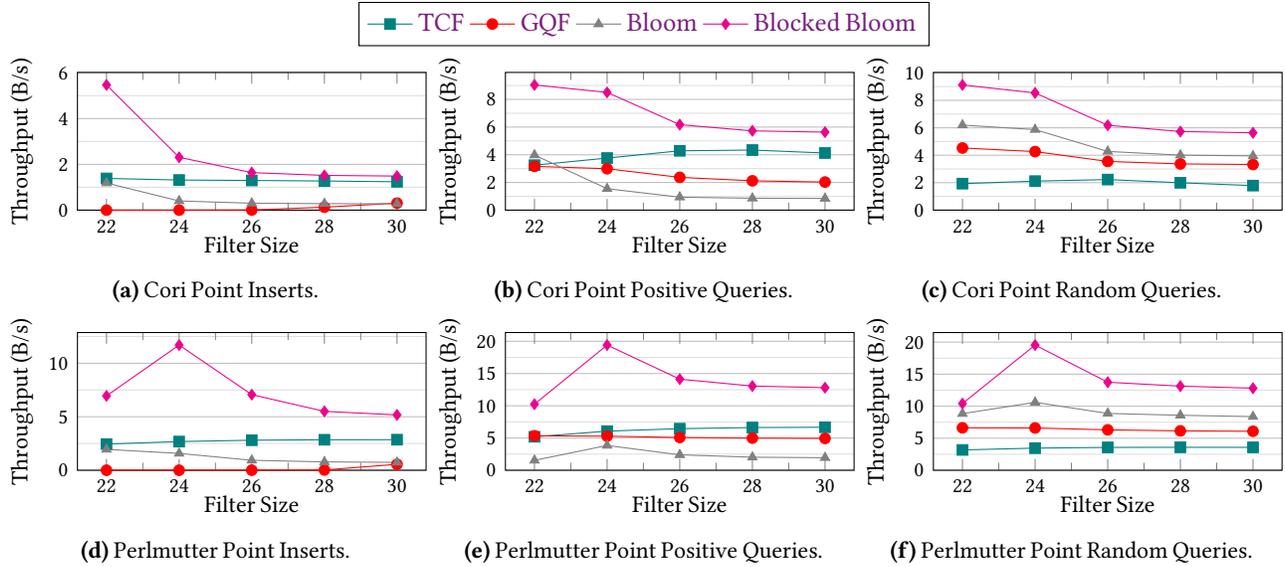
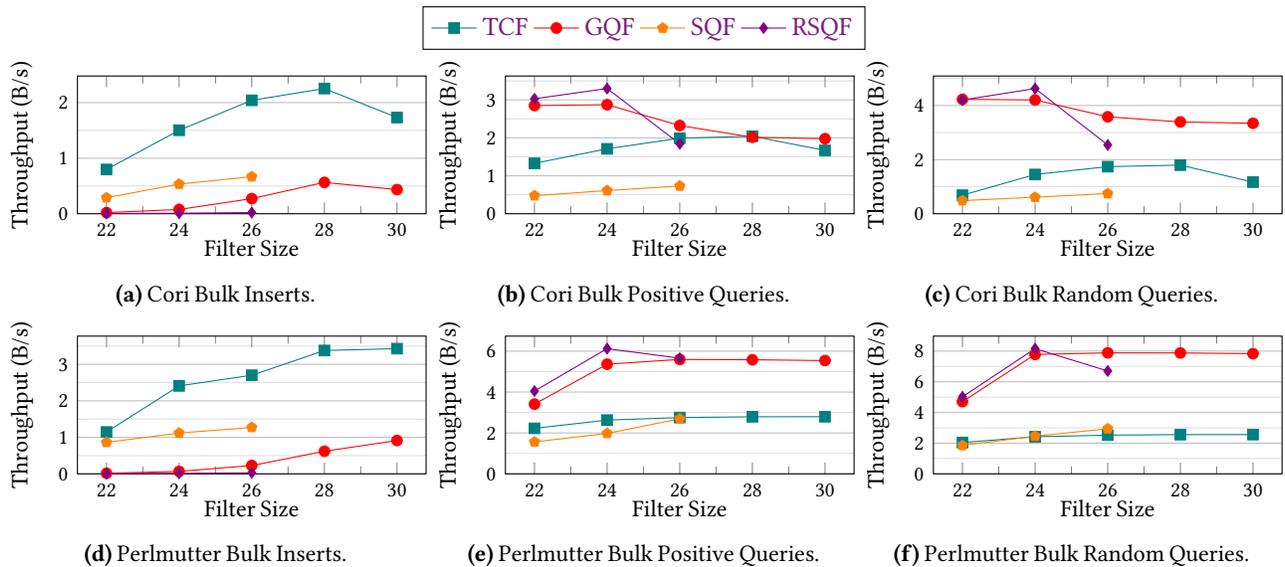
\begin{figure*}
   \centering
   \ref{BulkAgg}\\
   \begin{subfigure}{0.32\linewidth}
      \begin{tikzpicture}
         \begin{axis}[
               AggregatePlot,
               legend entries={TCF, GQF, SQF, RSQF},
               legend columns=4,
               legend to name={BulkAgg}
            ]
            \addplot[TCQFStyle] table[x=x_0, y=y_0] {data/bulk_tcqf/bulk_tcqf_agg_insert.txt};
            \addplot[GQFStyle] table[x=x_0, y=y_0] {data/gqf/gqf_agg_insert.txt};
            \addplot[SQFStyle] table[x=x_0, y=y_0] {data/bulk_sqf/bulk_sqf_agg_insert.txt};
            \addplot[RSQFStyle] table[x=x_0, y=y_0] {data/rsqf/rsqf_agg_insert.txt};
         \end{axis}
      \end{tikzpicture}
      \caption{Cori Bulk Inserts.}
      \label{bulk-inserts}
   \end{subfigure}%
   \begin{subfigure}{0.32\linewidth}
      \begin{tikzpicture}
         \begin{axis}[
               AggregatePlot
            ]
            \addplot[TCQFStyle] table[x=x_0, y=y_0] {data/bulk_tcqf/bulk_tcqf_agg_lookup.txt};
            \addplot[GQFStyle] table[x=x_0, y=y_0] {data/gqf/gqf_agg_lookup.txt};
            \addplot[SQFStyle] table[x=x_0, y=y_0] {data/bulk_sqf/bulk_sqf_agg_lookup.txt};
            \addplot[RSQFStyle] table[x=x_0, y=y_0] {data/rsqf/rsqf_agg_lookup.txt};
         \end{axis}
      \end{tikzpicture}
      \caption{Cori Bulk Positive Queries.}
      \label{bulk-exists}
   \end{subfigure}%
   \begin{subfigure}{0.32\linewidth}
      \begin{tikzpicture}
         \begin{axis}[
              AggregatePlot
            ]
            
            \addplot[TCQFStyle] table[x=x_0, y=y_0] {data/bulk_tcqf/bulk_tcqf_agg_fp.txt};
            \addplot[GQFStyle] table[x=x_0, y=y_0] {data/gqf/gqf_agg_fp.txt};
            \addplot[SQFStyle] table[x=x_0, y=y_0] {data/bulk_sqf/bulk_sqf_agg_fp.txt};
            \addplot[RSQFStyle] table[x=x_0, y=y_0] {data/rsqf/rsqf_agg_fp.txt};
            
         \end{axis}
      \end{tikzpicture}
      \caption{Cori Bulk Random Queries.}
      \label{bulk-false}
   \end{subfigure}%
   \\
      \begin{subfigure}{0.32\linewidth}
      \begin{tikzpicture}
         \begin{axis}[
               AggregatePlot
            ]
            \addplot[TCQFStyle] table[x=x_0, y=y_0] {perlmutter_data/tcqf_bulk/insert.txt};
            \addplot[GQFStyle] table[x=x_0, y=y_0] {perlmutter_data/gqf/gqf-aggregate-insert.txt};
            \addplot[SQFStyle] table[x=x_0, y=y_0] {perlmutter_data/sqf/sqf-aggregate-insert.txt};
            \addplot[RSQFStyle] table[x=x_0, y=y_0] {perlmutter_data/rsqf/rsqf-aggregate-insert.txt};
         \end{axis}
      \end{tikzpicture}
      \caption{Perlmutter Bulk Inserts.}
      \label{perl-bulk-inserts}
   \end{subfigure}%
   \begin{subfigure}{0.32\linewidth}
      \begin{tikzpicture}
         \begin{axis}[
               AggregatePlot
            ]
            \addplot[TCQFStyle] table[x=x_0, y=y_0] {perlmutter_data/tcqf_bulk/lookup.txt};
            \addplot[GQFStyle] table[x=x_0, y=y_0] {perlmutter_data/gqf/gqf-aggregate-exists-lookup.txt};
            \addplot[SQFStyle] table[x=x_0, y=y_0] {perlmutter_data/sqf/sqf-aggregate-exists-lookup.txt};
            \addplot[RSQFStyle] table[x=x_0, y=y_0] {perlmutter_data/rsqf/rsqf-aggregate-exists-lookup.txt};
         \end{axis}
      \end{tikzpicture}
      \caption{Perlmutter Bulk Positive Queries.}
      \label{perl-bulk-exists}
   \end{subfigure}%
   \begin{subfigure}{0.32\linewidth}
      \begin{tikzpicture}
         \begin{axis}[
              AggregatePlot
            ]
            
            \addplot[TCQFStyle] table[x=x_0, y=y_0] {perlmutter_data/tcqf_bulk/fp.txt};
            \addplot[GQFStyle] table[x=x_0, y=y_0] {perlmutter_data/gqf/gqf-aggregate-false-lookup.txt};
            \addplot[SQFStyle] table[x=x_0, y=y_0] {perlmutter_data/sqf/sqf-aggregate-false-lookup.txt};
            \addplot[RSQFStyle] table[x=x_0, y=y_0] {perlmutter_data/rsqf/rsqf-aggregate-false-lookup.txt};
            
         \end{axis}
      \end{tikzpicture}
      \caption{Perlmutter Bulk Random Queries.}
      \label{perl-bulk-false}
   \end{subfigure}%
    \vspace{-0.5em}   
\caption{Bulk API aggregate throughput comparison between various filters with one batch.}
\label{fig:bulk-all-perf}
\end{figure*}
In this section, we evaluate the performance of various GPU filter implementations.
We compare our implementations of the two-choice filter (TCF) and GPU-based counting quotient filter (GQF) against Geil et al.'s~\cite{GeilFO18} standard quotient filter (SQF) and rank-select quotient filter (RSQF). The SQF is a GPU implementation of the quotient filter and supports insertions, queries, and deletions. The RSQF does not supports deletions. Both SQF and RSQF do not support counting. We configure the SQF and RSQF to achieve the best performance based on author's recommendations. 

As a baseline for the performance of a filter that does not support deletions, we also include the Bloom filter (BF) and blocked Bloom filter (BBF) in our evaluation.
The BF and BBF are not directly comparable to other filters used in the evaluation as they do not support similar features.
The BBF is taken from Junger et al.~\cite{warpcore} and is configured according to the author's recommendation to achieve best performance. We modified a C++ BF implementation~\cite{Bloom-code} to a 1-bit encoded GPU implementation using CUDA atomic bitwise operations.

%

\begin{table}[t]
\centering
\resizebox{\columnwidth}{!}{%
    \begin{tabular}{c | c c | c c | c c | c c }
    \toprule
    Filter & 
    \multicolumn{2}{c}{{\bf Insert}} & \multicolumn{2}{c}{{\bf Query}} & \multicolumn{2}{c}{{\bf Delete}} & \multicolumn{2}{c}{{\bf Count}} \\
    \midrule
    & Point & Bulk  & Point & Bulk & Point & Bulk & Point & Bulk \\
    \midrule
    GQF & \checkmark & \checkmark  & \checkmark & \checkmark  & \checkmark & \checkmark  & \checkmark & \checkmark \\ 
    TCF & \checkmark & \checkmark & \checkmark & \checkmark & \checkmark & \checkmark & & \\
    BF & \checkmark & \checkmark  & \checkmark & \checkmark  &   &    &   &   \\ 
    SQF &   & \checkmark  &   & \checkmark  &   & \checkmark &   &   \\ 
    RSQF &   & \checkmark  &   & \checkmark  &   &    &   &   \\ 
    \bottomrule
    \end{tabular}
    }
    \caption{API supported by various filters. The GQF is the only filter that supports a range of operations. RSQF can support deletes but it is not implemented by the authors.}
    \label{tab:api_table_checkmark}
    \vspace{-1.5em}
\end{table}

We evaluate each filter on two fundamental operations: insertions and lookups. Lookups are evaluated both for items that are present and for items that are not present in the filter.
Our evaluation of filters is split on the status of the filter as either \emph{bulk} or \emph{point} API. Point filters have device-side APIs and can be called to insert or query a single item while bulk filters must be called from a host function.
The TCF and GQF support both bulk and point APIs. We compare our bulk implementation of the TCF and GQF with the SQF and RSQF as they both are designed for bulk API. We compare our point implementations of the TCF and GQF with the Bloom filter and blocked Bloom filter. Both the Bloom and blocked Bloom implementations only support point API.

Please refer to~\Cref{tab:api_table_checkmark} for a complete list of API supported by various filters. Only the GQF and TCF support both bulk and point modes for insert, query, delete, and count operations. 
We compare the GQF and TCF only against the SQF for delete operations. The GQF is compared against no other filter for counting as no other filter supports counting.

\ppoppaddition{We also evaluate the memory reduction in MetaHipMer when using the TCF to filter singleton $k$-mers.}

\para{Microbenchmarks setup}
Our evaluation setup includes all the micro benchmarks employed by filter data structure papers~\cite{EinzigerFr16,BenderFaJo12a,PandeyBJP17,pandeySigmod21,FanAnKa14,BreslowJ18,Geil16,GeilFO18} in the past.

We measure performance on raw inserts and lookups as follows. We generate 64-bit input items from the hashed output of a cuRand XORWOW generator. Items are inserted into an empty filter until it reaches its maximum recommended load factor (e.g., 90\%).
For successful lookups, we query items that are already inserted. For random lookups, we generate a different set of 64-bit hashes than the set used for insertion. This is done by using the hashed outputs of an XORWOW generator set with a different seed.
We report aggregate throughput of the operations to insert or query a set of items.
%

One challenge that we face in designing our experiments is that the filters do not all support the same false-positive rate. For example, the GQF supports 8, 16, 32, and 64 bit remainders in order to keep the slots in the table machine-word aligned. This helps simplify the GPU implementation by avoiding memory conflicts when multiple threads are modifying different slots.
However, SQF and RSQF filters only support remainder sizes of 5 and 13 as they pack the 3 metadata bits along with the remainder in 8 and 16 bit machine words. They further require the sum of the quotient and remainder bits to be less then 32. Therefore, they can only support up to $2^{26}$ items with 5-bit remainders and $2^{18}$ items with 13-bit remainders.

\begin{table}[t]
\centering
\resizebox{\columnwidth}{!}{%
    \begin{tabular}{ c c | c c | c c | c c | c c | c c | c c}
    \toprule
    \multicolumn{2}{c}{{\bf GQF}} & \multicolumn{2}{c}{{\bf BF}} & \multicolumn{2}{c}{{\bf SQF}} & \multicolumn{2}{c}{{\bf RSQF}} &
    \multicolumn{2}{c}{{\bf Bulk TCQF}} &
    \multicolumn{2}{c}{{\bf TCQF}} &
    \multicolumn{2}{c}{\bf Blocked Bloom}\\
    \midrule
    FP & BPI  & FP & BPI & FP & BPI & FP & BPI & FP & BPI & FP & BPI & FP & BPI\\
    \midrule
    0.19\% & 10.68 & 0.15\% & 10.10 & 1.17\% & 9.7 & 1.55\% & 7.87 & 0.36\% & 16 & .024\% & 16 & .71\% & 9.73 \\
    \bottomrule
    \end{tabular}
    }
    \caption{False-positive rate (FP) and bits per item (BPI) of various filters for experiments in~\Cref{fig:bulk-all-perf} and \Cref{fig:point-performance}. } 
    \label{tab:merged_fp_space}
    \vspace{-2.0em}
\end{table}

We pick a target false-positive rate of .1\% and configure each filter to get as close to this false positive rate as possible. We use 8-bit remainders in the GQF. We use 7 hashes and ~10.1 bits per item in the Bloom and blocked Bloom filter. We use 5-bit remainders for the SQF and RSQF and although this results in almost an order-of-magnitude higher false-positive rates, it supports the largest number of items ($2^{26}$) for these implementations. The smallest TCF word alignment under this error rate is 16 bits, so we report the results from this variation of the filter. Table \ref{tab:merged_fp_space} shows the empirical space usage, false-positive rate, and bits-per-item (BPI) of different filters in these experiments. We measure the space-usage and false-positive rates empirically.


We evaluate the performance of these filters in the GPU memory and hence we size the filters in our experiments so that they can always reside in the GPU memory.

\para{Counting benchmark setup}
The counting benchmarks include three datasets with different count distributions. The uniform-random (UR) dataset contains items drawn from a uniform-random distribution with almost no duplicates. The uniform-random count (UR count) dataset contains items where the counts of items are drawn from a uniform-random distribution between 1 and 100. The zipfian count (Zipfian count) dataset contains items where the counts of items are drawn from a Zipfian distribution (the coefficient is 1.5 and items are chosen from a universe of the same size as the dataset). All the items in the dataset are inserted in one big batch in the GQF.
We also include a real-world genomic dataset for the counting benchmark. We took a raw sequencing file, \emph{M. balbisiana}, from the Squeakr~\cite{PandeyBJP17b} benchmark dataset and extracted $k$-mers for counting. 

\begin{figure*}
   \centering
   \ref{cgperflegend}\\
   \begin{subfigure}{0.32\linewidth}
      \begin{tikzpicture}
         \begin{axis}[
               CGPlot,
               legend entries={8-8, 12-8, 12-12, 12-16, 12-32, 16-16, 16-32},
               legend columns=7,
               legend to name={cgperflegend},
               xmode=log,
               log basis x ={2}
            ]
            \addplot[88Style] table[x=cg, y=insert] {data/3D_data/8_8.txt};
            \addplot[128Style] table[x=cg, y=insert] {data/3D_data/12_8.txt};
            \addplot[1212Style] table[x=cg, y=insert] {data/3D_data/12_12.txt};
            \addplot[1216Style] table[x=cg, y=insert] {data/3D_data/12_16.txt};
            \addplot[1232Style] table[x=cg, y=insert] {data/3D_data/12_32.txt};
            \addplot[1616Style] table[x=cg, y=insert] {data/3D_data/16_16.txt};
            \addplot[1632Style] table[x=cg, y=insert] {data/3D_data/16_32.txt};
           
         \end{axis}
      \end{tikzpicture}
      \caption{Inserts.}
      \label{cginsert}
   \end{subfigure}%
   \begin{subfigure}{0.32\linewidth}
      \begin{tikzpicture}
         \begin{axis}[
               CGPlot,
               xmode=log,
               log basis x ={2}
            ]
            \addplot[88Style] table[x=cg, y=query] {data/3D_data/8_8.txt};
            \addplot[128Style] table[x=cg, y=query] {data/3D_data/12_8.txt};
            \addplot[1212Style] table[x=cg, y=query] {data/3D_data/12_12.txt};
            \addplot[1216Style] table[x=cg, y=query] {data/3D_data/12_16.txt};
            \addplot[1232Style] table[x=cg, y=query] {data/3D_data/12_32.txt};
            \addplot[1616Style] table[x=cg, y=query] {data/3D_data/16_16.txt};
            \addplot[1632Style] table[x=cg, y=query] {data/3D_data/16_32.txt};
            
         \end{axis}
      \end{tikzpicture}
      \caption{Positive queries.}
      \label{cglookup}
   \end{subfigure}%
   \begin{subfigure}{0.32\linewidth}
      \begin{tikzpicture}
         \begin{axis}[
               CGPlot,
               xmode=log,
               log basis x ={2}
            ]
            
            \addplot[88Style] table[x=cg, y=fp] {data/3D_data/8_8.txt};
            \addplot[128Style] table[x=cg, y=fp] {data/3D_data/12_8.txt};
            \addplot[1212Style] table[x=cg, y=fp] {data/3D_data/12_12.txt};
            \addplot[1216Style] table[x=cg, y=fp] {data/3D_data/12_16.txt};
            \addplot[1232Style] table[x=cg, y=fp] {data/3D_data/12_32.txt};
            \addplot[1616Style] table[x=cg, y=fp] {data/3D_data/16_16.txt};
            \addplot[1632Style] table[x=cg, y=fp] {data/3D_data/16_32.txt};
            
         \end{axis}
      \end{tikzpicture}
      \caption{Random Queries.}
      \label{cgfp}
   \end{subfigure}%
   \vspace{-1.0em}
\caption{Comparison between cooperative group sizes. All tests were run on filters sized to $2^{28}$. The left number in a label is fingerprint size and the right is the block size.}
\label{fig:cgcomparison}
\vspace{-1.0em}
\end{figure*}
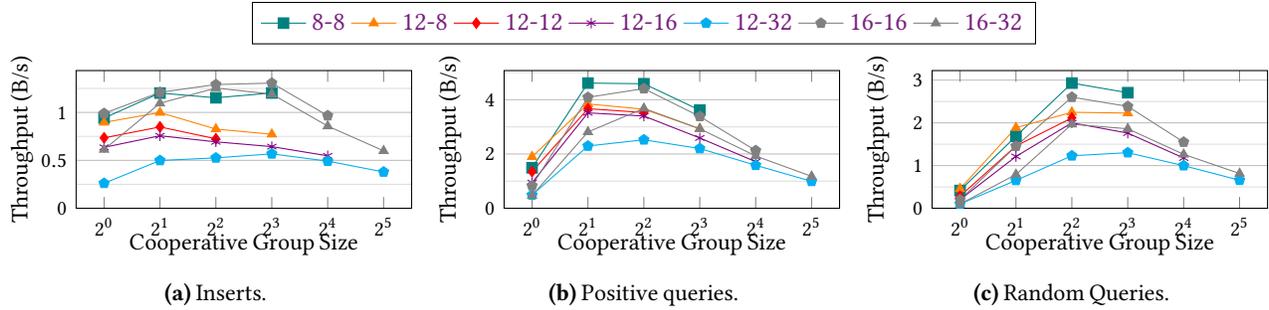
\para{Machine specification}
Our microbenchmarks and counting benchmarks were run on Cori’s~\cite{cori}  and Perlmutter's~\cite{perlmutter} GPU nodes. Cori nodes consists of NVIDIA Tesla V100 with 5120@1445MHz microprocessors, 16 GB 4096-bit HBM2 memory, and an active thread limit of 82,000 simultaneous threads. Perlmutter nodes consists of NVIDIA A100 Tensor Core GPU with 6912@1410MHz 40 GB 5120-bit HBM2 memory and an active thread limit of 110,000 threads.


\subsection{Point API Performance}

The results of the point API benchmarking an be found in \Cref{fig:point-performance}. The TCF has the highest insert and query performance among the filters that support insertion, queries, and deletions. 
It requires two cache line probes for most queries and one write for insertions which is much smaller than all other filters.

The overhead of the backing table is negligible as less than 0.07\% of items go in the backing table. However, for negative queries (i.e., the items not present in the filter), the backing table adds to the worst-case performance:
the query must check at least one bucket in the backing table, and can probe up to 20 buckets in the worst case. 
\ppoppaddition{The backing table helps achieve 90\% load factor. Without the backing table the TCF could only get to 79.6\% load factor before failing to insert an item.}
Furthermore, the average performance of insert and query operations is much better due to the shortcut optimization mentioned in \Cref{sec:TCF_impl}.

The TCF has a higher ($\approx2\times$) false-positive rate compared to the GQF and BF in this evaluation. However, the TCF supports multiple configurations in terms of the space usage and false-positive rate. We have evaluated the performance of various TCF configurations in~\Cref{sec:tcf-variations}.


The GQF performance is slower compared to the TCF due to the overhead of locking to perform point insertions. The locking implementation requires us to maintain separate locks for each chunk in the GQF and this causes lock thrashing.
Based on the positive query performance, the GQF can reach a slot for insertion faster than the BF can operate on all 7 bits, as each bit requires a different cache load in the BF. However, the cost of locking is so prohibitive on GPUs that the BF is faster for insertions as all operations occur without thrashing.


\begin{figure}
   \centering
   \ref{agg-delete-legend-load}\\
      \begin{tikzpicture}
         \begin{axis}[
               ymode=log,
               log ticks with fixed point,
               DeletePlot,
               legend entries={\color{black}GQF-Bulk,  \color{black}SQF, \color{black}TCF},
               legend columns=3,
               legend to name={agg-delete-legend-load}
            ]
          \addplot[GQFStyle]  table[x=x_0, y=y_0] {data/deletes/gqf_deletes.txt};
          \addplot[SQFStyle]   table[x=x_0, y=y_0] {data/deletes/sqf_deletes.txt};
          \addplot[TCQFStyle]   table[x=x_0, y=y_0] {data/deletes/tcqf_deletes.txt};
         \end{axis}
      \end{tikzpicture}
      \vspace{-0.5em}
      \caption{Deletion performance of GQF bulk, SQF, and TCF on Cori GPU nodes. The x-axis shows $\log{n}$, where $n$ is the number of slots in the filter. SQF only support up to $2^{26}$ slots.}
      \label{fig:agg-deletes}
\end{figure}
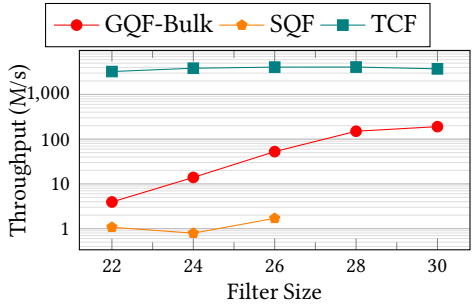
\paragraph{Bloom filters}

The BBF is the faster of the two filters. It requires a single cache line operation and uses atomicOR which is faster than atomicCAS required by other filters.
However, the BBF has $\approx5.5\times$ higher false positive rate when compared to a Bloom filter with the same bits per item (~\Cref{tab:merged_fp_space}).



The Bloom filter has relatively low throughput on inserts and random queries, as it needs to check multiple random slots within the filter, each of which requires a different cache line load. 
The BF shows relatively higher throughput for random lookups, as it has a high probability of finding a zero and terminating the search early.

The BF and BBF have outlier performance at $2^{22}$ for Cori and $2^{24}$ for Perlmutter. This is due to these filters being small enough to fit within the L2 cache, allowing for faster memory operations and saturating all the GPU threads efficiently. 

\subsection{Bulk API performance}

The results of the bulk API benchmarking can be found in \Cref{fig:bulk-all-perf}. The bulk TCF is the fastest filter for inserts, with a maximum throughput of over 3.4 Billion per second on Perlmutter. However, as this filter relies on binary search to find items within a bucket, it has lower throughput on queries, topping out at $\approx2$ Billion per second for both positive and random queries. The bulk SQF has the next highest insert throughput, though the sorted bulk lookup strategy used in the SQF has lower throughput than the other filters. 

The throughput of the bulk GQF depends on the size of the filter, so we see an increase in performance as the filter grows, topping at $2^{28}$ when the parallelism in the GPU is saturated. The queries in the bulk GQF scale directly with the number of items, so the positive and random queries show high performance even on small filter sizes.

The RSQF has very high throughput on both types of queries. The performance drops as the filter grows to $2^{26}$, due to the filter exceeding the 8 MB size of the V100 L2 cache. The filter has very poor performance on inserts, topping out at 8 Million per second, roughly three orders of magnitude lower than the other filters. As the RSQF and GQF have very similar internals, there is no reason the inserts of the filter could not be accelerated. However, an optimized function for inserts is the provided by the authors.

For inserts, all of the bulk filters show increasing throughput with dataset size. The insert schemes used in these filters map CUDA threads or warps to sections of memory. This results in far less active threads than the point filters, which map warps to individual items and can quickly reach saturation.

\subsection{TCF variations}\label{sec:tcf-variations}

\Cref{fig:cgcomparison} shows the performance effects of modulating the cooperative group size for a variety of TCF filter variations. These results show that there is an optimal cooperative group sizing for each filter variation. For the majority of the configurations, this size is 4. These optimal sizes are an effect of the trade off between compute and memory latency due to how warps, and by extension cooperative groups, are scheduled on streaming multiprocessor.

Shrinking the number cooperative groups increases the saturation of the memory pipeline while lowering the amount of compute available per cooperative group. 
Increasing the size of the cooperative group gives less divergence and better compute throughput at the expense of less memory operations being scheduled.
When memory and compute are balanced, the filter can entirely overlap computation and communication, leading to the most efficient performance.  For most designs, this optimal point occurs at a cooperative group of size 4, though some of the larger bucket designs also perform well at 8 due to the extra work to traverse a bucket.

The 8 and 16 bit versions of the filter have the fastest performance, as inserts and queries can be performed in one transaction. As 50\% of operations require two memory transactions, the 12 bit filters are slower than their counterparts.

\subsection{Deletion performance}
~\Cref{fig:agg-deletes} shows the performance for deletions for filters that support the operation. 
The TCF is an order-of-magnitude faster for deletes than the GQF, as the filter deletes items by replacing them with a dedicated tombstone key. This means that deletions can be done with one atomicCAS operation.
The GQF is up to two orders of magnitude faster for deletion than the SQF. This is due to the even-odd phased approach that minimizes the amount of left shifting that is required during a delete operation. Left shifting is further reduced due the sorting of items before the operation and deleting larger items first. Overall deletes are slower compared to the inserts in the GQF as deletes are more compute intensive.

\begin{table}[t]
\label{Tab:Merged}
\centering
\resizebox{\columnwidth}{!}{%
    \begin{tabular}{c c c c c | c }
    \toprule
    Dataset & Method & Nodes & TCF mem & HT mem & Total Mem \\
    \midrule
   \multirow{2}{*}{{\bf WA}} & TCF & 64 & 13 & 594 & 607 \\
    & No TCF & 64 & 0 & 1742 & 1742 \\
    \midrule

    \multirow{2}{*}{{\bf Rhizo}} & TCF & 64 & 27 & 119 & 146 \\
    & No TCF & 64 & 0 & 790 & 790 \\

    \bottomrule
    \end{tabular}
    }
    \caption{TCF and Hash table memory usage during MetaHipMer runs across three datasets. {\bf Memory is aggregated across all nodes and is in GB.}}
    \label{tab:mhm_tcf}
    \vspace{-1em}
\end{table}

\subsection{MetaHipMer Performance}

\ppoppaddition{
We integrated the TCF into the $k$-mer analysis phase of MetaHipMer (MHM)~\cite{GeorganasEHG18,HofmeyrEGC20}. MHM uses GPUs to accelerate $k$-mer counting which is the most memory intensive phase in the pipeline.
The TCF helps to weed out singleon $k$-mers which can take up to 70\% of the memory. 
Using the TCF reduces the memory usage of the entire application by 38\%. The effect of the TCF on two assembly of genomic datasets, a 813GB sample from the Western Artic Ocean (WA) and a 129GB dataset of biofuel crop reads (Rhizo), can be seen in \cref{tab:mhm_tcf}.
}

\begin{table}[t]
\centering
\resizebox{\columnwidth}{!}{%
    \begin{tabular}{ c || c | c | c | c }
    \toprule
    Filter & Size & Inserts & Positive Queries & Random Queries\\
    & & per Second & per Second & per Second \\
    \midrule
    CQF & 28 & 2.2  & 320.9  & 368.0 \\
    \midrule 
    Point GQF & 28 & 129.7 & 2118.4  & 3369.0 \\
    \midrule 
    \midrule
    VQF & 28 & 247.2  & 332.0 & 333.8\\
    \midrule 
    Point TCF & 28 & 1273.8 & 4340.9 & 1994.3 \\
    \bottomrule
    \end{tabular}
    }
    \caption{\ppoppaddition{Aggregrate throughput for the CPU and GPU versions of the filters. }}
    \label{tab:cpugputable}
\end{table}

\subsection{CPU Performance}

\ppoppaddition{
\Cref{tab:cpugputable} shows the difference in performance between the CPU filters (CQF~\cite{PandeyBJ17} and VQF~\cite{pandeySigmod21}) and the corresponding GPU filter designs. The CPU filters were run on Cori's KNL nodes with 272 threads per filter. The GPU filters were run on Cori's GPU nodes. All filters were benchmarked for throughput on inserts, queries, and random queries. The GPU filters are up to two order-of-magnitude faster than their CPU counter parts. 
The speed ups achieved by the GQF and TCF are significantly higher than previous GPU filters, the SQF and RSQF~\cite{GeilFO18}. This shows the advantages of this new GPU-optimized filer design.
}

\begin{table}[t]
\centering
\resizebox{\columnwidth}{!}{%
    \begin{tabular}{ c || c | c | c | c | c}
    \toprule
    Size & UR & UR count & Zipfian count & Zipfian Count (MR) & $k$-mer count \\
    \midrule
    22 & 25.318  & 30.763  & 3.676 & 34.888 & 23.625 \\
    \midrule          
    24 & 101.804 & 110.833 & 4.777 & 169.637 & 90.722 \\
    \midrule          
    26 & 321.150 & 350.824 & 4.995 & 508.156 & 296.130 \\
    \midrule          
    28 & 566.038 & 798.353 & 4.520 & 806.766 & 507.373 \\
    \bottomrule
    \end{tabular}
    }
    \caption{Aggregate insertion throughput of GQF (Million operations/sec) for inserting (counting) items from datasets with different distributions. 
    Zipfian count (MR): count of items are drawn from a Zipfian distribution using the Map-reduce implementation from~\cref{sec:skew}.}
    \label{tab:count-gqf}
\end{table}

\subsection{Counting performance in the GQF}
\Cref{tab:count-gqf} shows the aggregate insertion throughput for inserting (counting) items from datasets with three different distributions. Counting items from a Zipfian distribution using the map-reduce (Zipfian MR) strategy explained in~\Cref{sec:skew} achieves the highest throughput.



When counting items, especially when the counts are smaller than the maximum value in a GQF slot (which is 256 for a 8-bit slot), the insertions mostly involve incrementing the count of an existing item, which can be done fairly efficiently without the need to shift remainders. However, when the distribution is skewed, as in the case of a Zipfian distribution, many threads contend to insert the same item, causing long stalls which reduce throughput. This shows that the GQF is an efficient counting filter for datasets with small counts.

For the $k$-mer counting dataset, the GQF supports throughput of more than 500M $k$-mers per second which is orders of magnitude faster than the throughput of Squeakr~\cite{PandeyBJP17b}, a CPU $k$-mer counter built using the CQF. With the GQF, we can easily port Squeakr to GPUs and accelerate $k$-mer counting.



\subsection{Discussion}

For most data analytics applications, the TCF is appropriate the choice for a GPU filter. \ppoppaddition{This is in large part due to the stability of the filter which makes it resilient to skew and under-sizing.} It offers the right trade off between space efficiency and false positive rate, maintains high throughput for all operations, scales to larger datasets, and can be configured for a wide range of filtering use cases.
For applications that require no associativity and are not bound by space usage or false positive rate, the blocked Bloom filter (BBF) is a good choice.
The rich features of the GQF are critical to many analytics applications like MetaHipMer, database merges, etc. However, this comes at an additional performance cost.
The GQF is often the only available filter option for many applications that need GPUs to accelerate complex data processing.

\section*{Acknowledgments}

This research is funded in part by
the Advanced Scientific Computing Research (ASCR)
program within the Office of Science of the DOE under contract number
DE-AC02-05CH11231,
the Exascale Computing Project (17-SC-20-SC), a collaborative
effort of the U.S. Department of Energy Office of Science and the National
Nuclear Security Administration.
We used resources of the NERSC supported by the Office of
Science of the DOE under Contract No. DEAC02-05CH11231.


\balance
%
\bibliographystyle{ACM-Reference-Format}

\typeout{}
\bibliography{bibliography, allpapers, BFJ-bigdata-papers}  
\clearpage

\end{document}